# Measuring Hall Viscosity of Graphene's Electron Fluid

A. I. Berdyugin[1], S. G. Xu [1,2], F. M. D. Pellegrino[3,4], R. Krishna Kumar[1,2], A. Principi[1], I. Torre[5], M. Ben Shalom[1,2], T. Taniguchi[6], K. Watanabe[6], I. V. Grigorieva[1], M. Polini[7,1], A. K. Geim[1,2], D. A. Bandurin[1]

[1]School of Physics & Astronomy, University of Manchester, Manchester M13 9PL, United Kingdom
[2]National Graphene Institute, University of Manchester, Manchester M13 9PL, United Kingdom
[3]Dipartimento di Fisica e Astronomia, Università di Catania, Via S. Sofia, 64, I-95123 Catania, Italy
[4]INFN, Sez. Catania, I-95123 Catania, Italy
[5]ICFO - Institut de Ciencies Fotoniques, The Barcelona Institute of Science and Technology, 08860 Castelldefels (Barcelona), Spain.
[6]National Institute for Materials Science, 1-1 Namiki, Tsukuba, 305-0044 Japan
[7]Istituto Italiano di Tecnologia, Graphene Labs, Via Morego 30, 16163 Genova, Italy

*Materials subjected to a magnetic field exhibit the Hall effect, a phenomenon studied and understood in fine detail. Here we report a qualitative breach of this classical behavior in electron systems with high viscosity. The viscous fluid in graphene is found to respond to non-quantizing magnetic fields by producing an electric field opposite to that generated by the classical Hall effect. The viscous contribution is large and identified by studying local voltages that arise in the vicinity of current-injecting contacts. We analyze the anomaly over a wide range of temperatures and carrier densities and extract the Hall viscosity, a dissipationless transport coefficient that was long identified theoretically but remained elusive in experiment. Good agreement with theory suggests further opportunities for studying electron magnetohydrodynamics.*

Despite decades-long interest and extensive theoretical efforts, studies of electron hydrodynamics have been hampered by the lack of suitable experimental systems (*1, 2*). The situation has changed recently due to availability of new materials (*3–6*) in which both impurity and electron-phonon scattering – that do not conserve the electron momentum – are relatively weak so that electron-electron collisions become the dominant source of scattering. In particular, graphene was shown to provide an excellent platform for studying electron hydrodynamics due to its high quality and exceptionally weak electron-phonon coupling (*7, 8*). The latter ensures that electron-electron scattering dominates over a wide range of temperatures ($T$), from liquid-nitrogen to room, and results in the electron kinematic viscosity $\nu$ as high as ~0.1 m$^2$ s$^{-1}$, exceeding that of honey (*3, 9*). Experiments in graphene have provided clear evidence for fluid-like behavior of electrons, which reveals itself in negative vicinity resistance (*3*), breakdown of the Wiedemann-Franz law (*4*) and superballistic transport (*9*). So far, studies of electron hydrodynamics focused on zero magnetic field $B$. Only recently, several theoretical reports suggested possible observables for magnetohydrodynamics such as large negative magnetoresistance (*10–13*) and anomalous Hall resistivity (*11, 12*).

Let us first get some qualitative insight into the magnetohydrodynamic behavior expected for two-dimensional electron fluids. To this end, Figs. 1a and b plot the electric potential distribution $\phi(\mathbf{r})$ expected near a narrow current injector in zero and finite $B$, respectively. The injected current $I$ entrains adjacent fluid regions, which results in negative lobes of the potential near the injector (*14, 15*). In zero $B$ (Fig. 1a), the lobes are symmetric with respect to the injection direction and, for restricted geometries, can be accompanied by whirlpools of electrical current (*14–18*). A finite $B$ induces a considerable asymmetry in $\phi(\mathbf{r})$ (Fig. 1b). This contains the following three contributions. First, the classical Hall effect (HE) causes the familiar potential difference $V_{\mathrm{H}} = IB/ne$ between the left and right sides of the half-plane (Fig. 1c) where $n$ is the carrier density and $e$ the elementary charge. The second contribution is due to the longitudinal viscosity $\nu(B)$ and, in small $B$, is practically indistinguishable from that shown in Fig. 1a for zero $B$. The third contribution arises from the Hall viscosity $\nu_{\mathrm{H}}$, an off-diagonal component of the viscosity tensor (*19, 20*), which is the subject of our interest in this report. The $\nu_{\mathrm{H}}$ contribution is shown in Fig. 1D and has the sign opposite to that in Fig. 1c. It is clear that $\nu_{\mathrm{H}}$ suppresses the normal Hall response but its influence rapidly decays away from the injector region (Fig. 1d). The latter behavior makes it difficult to observe Hall viscosity using conventional devices and measurement geometries. As shown below, the vicinity geometry (Fig. 1e) has allowed us not only to separate qualitatively the standard HE from the one caused by $\nu_{\mathrm{H}}$ but also to accurately measure it.

Our devices were multiterminal Hall bars such shown in Fig. 1e and figs. S1a-b. They were made from graphene encapsulated between hexagonal boron-nitride crystals using the standard dry-transfer procedures (*18*). The Hall bars had widths of $2-4$ μm and were endowed with narrow (~0.3 μm) and closely spaced (~0.5 μm) voltage probes (Fig. 1e). Such submicrometer probes are essential for detection of viscous effects as seen from the spatial scale of Figs. 1a-d. Several devices made from mono- and bi- layer graphene (MLG and BLG, respectively) were studied, and all exhibited similar behavior. They had typical mobilities exceeding $\sim 100{,}000$ cm$^2$ V$^{-1}$ s$^{-1}$ at all $T$ up to 300 K, which ensured micrometer-scale transport with respect to momentum-non-conserving scattering over the entire $T$ range used in the experiments (fig. S1).

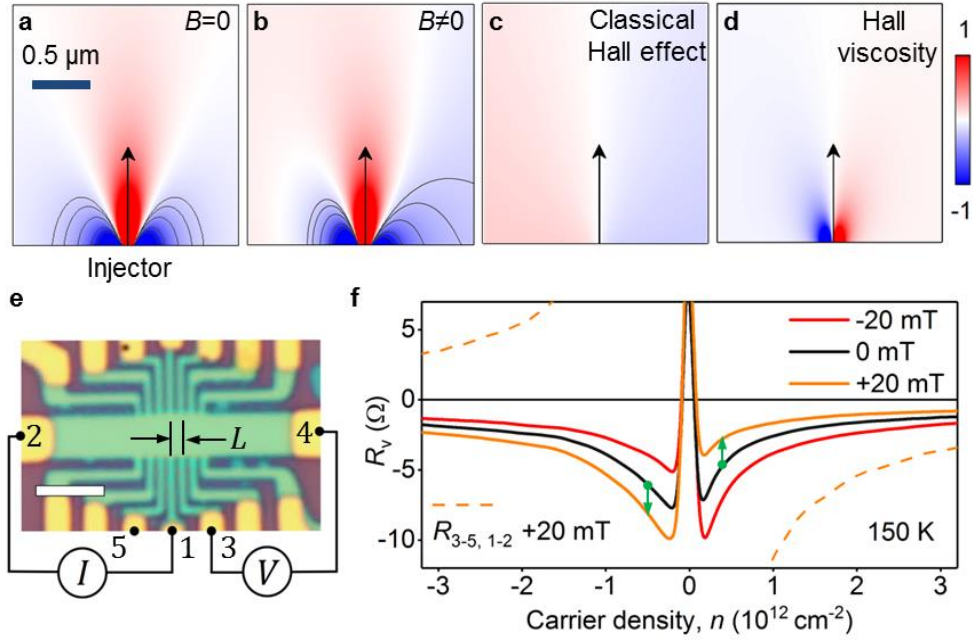

FIG. 1: Effect of magnetic field on viscous electron flow. (**a** and **b**) Electric potential distribution $\phi(\mathbf{r})$ expected in graphene's electron fluid near a current injector in zero $B$ and 50 mT, respectively. The calculations (*12*) use typical values of $\nu$ and $\nu_H$, found in our experiments below, and characteristic $n = 2 \times 10^{12}$ cm$^{-2}$ and $\tau = 2$ ps. Solid curves: Equipotentials. Box size, 2 μm × 2 μm. (**c**) Contribution to $\phi(\mathbf{r})$ from the classical Hall effect towards the map in (b). (**d**) Contribution that comes solely from $\nu_H$. (**e**) Optical micrograph of one of our devices, along with the schematic of the vicinity geometry. Scale bar, 3 μm. (**f**) Examples of the vicinity resistance for different $B$ (solid curves). $L \approx 1$ μm. Dashed: Local Hall resistance measured using voltage probes 3 and 5 close to the current injector (*18*).

In the vicinity geometry, the current $I$ is injected through a narrow contact (e.g., probe 1 in Fig. 1e) into a wide graphene channel, and a local potential $\phi$ is measured using probe 3 positioned at the distance $L$ from the injector. Contacts 2 and 4 (chosen sufficiently far away from the injection region) complete the electric circuit, serving as the drain and reference-voltage contacts, respectively. The vicinity resistance is defined as $R_v = R_{34,12} = V_{34}/I_{12}$ where $V_{34}$ is the voltage drop between 3 and 4. As per Figs. 1a-d, $R_v$ is expected to be sensitive to viscous effects if $L$ is sufficiently small (*3*, *14*, *15*, *18*). According to theory and previous experiments, the negative sign of $R_v$ can serve as a clear indicator of highly viscous flow (*3*, *14*, *15*, *18*). Indeed, in all our devices, $R_v$ is found negative above liquid-nitrogen $T$ and can remain negative up to room $T$ (fig. S2). At lower $T$, electron transport becomes dominated by single-particle ballistic effects that result in positive $R_v$ (fig. S2). For the purpose of this report, we focus on the $T$ range where the hydrodynamics dominates. In addition, we set several other constraints on variables used in our experiments. First, we limit ourselves to $B < 40$ mT such that the cyclotron radius always exceeds the width of our devices. This is to avoid obscuring effects caused by Landau quantization and electron focusing (*21*). Second, to avoid unrelated effects due to thermal excitations and charge inhomogeneity, we carry out experiments away from the charge neutrality point, at $n$ of the order of $10^{12}$ cm$^{-2}$. Finally, we employ small $I \leq 1$ µA to stay in the linear response regime and avoid nonlinear effects including electron heating (*2*, *3*).

Examples of $R_v(n)$ are shown in Fig. 1f using one of our MLG devices at 150 K where $R_v$ becomes most negative indicating strong viscous contributions (*3*). In zero $B$, $R_v$ is negative for all $n$ away from the charge neutrality and is practically symmetric for electron and hole doping (positive and negative $n$, respectively). A small positive field of 20 mT shifts the $R_v$ curves in the opposite directions for electrons and holes, as indicated by the green arrows in Fig. 1f. The shifts are opposite for negative $B$ (red curve). This behavior implies a contribution that is antisymmetric with respect to $B$ and $n$, similar to the Hall effect. However, the classical HE cannot possibly explain the observed shifts because in the vicinity geometry voltage probes are placed on the same side of the current path, which cancels the HE contribution to the measured voltages. A formal proof of this can be found in Ref. (*12*). Experimentally, we have also checked that there is no classical HE contribution in the vicinity geometry using similar devices but with a low mobility. Furthermore, it is important to compare the sign of the vicinity-voltage changes induced by $B$ with the sign of the classical HE. To keep the same sign convention for $B$ and $n$, it is instructive to measure the local Hall resistance $R_{35,12}$ (Fig. 1e and fig. S6) instead of using the standard Hall geometry. In this case, we use contact 5 instead of 4 and keep all the other contacts same as in the $R_v$ measurements. This swap places the voltage probes at the opposite sides of the current path, giving rise to the voltage drop $V_H$ due to the classical HE. The antisymmetric-in-$B$ part of $R_{35,12}$ (to avoid a contribution from longitudinal resistivity) is plotted in Fig. 1f. It shows that the classical HE induces $\phi$ of the opposite polarity with respect to those causing the $B$-shifts in $R_v$. Indeed, the vicinity curve in Fig. 1f is shifted, for example, upwards for hole doping and positive $B$, whereas the classical HE would shift it downwards [see, also, section 6 of (*18*)]. This behavior agrees well with the opposite signs of the contributions expected from $V_H$ and $\nu_H$ towards $R_v$ as shown in Figs. 1c and d. This is our main qualitative observation and, as argued below, it is caused by the Hall viscosity.

For further analysis, we define the Hall (odd) component of the vicinity resistance as $R_A(B) = [R_v(B) - R_v(-B)]/2$. The antisymmetrization removes the contributions that are symmetric in $B$ and caused by the longitudinal viscosity $\nu$ and Ohmic flow (*3*, *12*). Fig. 2a and fig. S3a show examples of the $R_A(B)$ curves for a MLG and BLG devices, respectively. Within the studied interval of $T$ and for small $B$, the dependences are linear in $B$ for all the measured devices and for all $L$. By analogy with the conventional Hall coefficient, $\alpha_{\mathrm{H}} = R_{\mathrm{H}} ne/B \equiv 1$, it is instructive to introduce the viscous Hall coefficient, $\alpha_{\mathrm{VH}} = R_{\mathrm{A}} ne/B$ (*18*). In this form, the antisymmetric contribution $R_{\mathrm{A}}$ is effectively normalized by the classical HE, which provides a good sense of the magnitude for viscous effects.

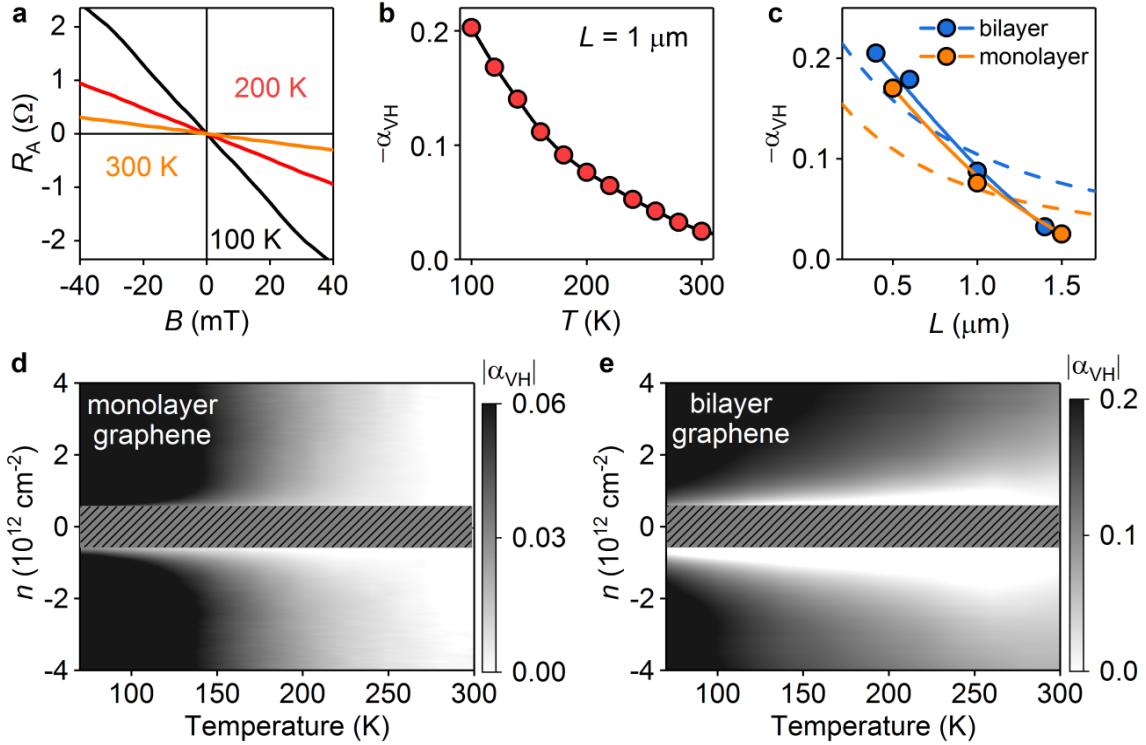

FIG. 2: Viscous Hall effect. (**a**) $R_{\mathrm{A}}(B)$ for one of our MLG devices; $L \approx 1$ μm. (**b**) The dimensionless viscous coefficient $\alpha_{\mathrm{VH}}$ as a function of $T$ (symbols). (**c**) $\alpha_{\mathrm{VH}}$ as a function of $L$ at 200 K (symbols). Dashed lines: Calculated dependences using Eq. (2) with $\nu_0$ from Ref. (*3*). $n = 2 \times 10^{12}$ cm⁻² for (a-c). Solid curves in (b,c): Guides to the eye. (**d** and **e**) $|\alpha_{\mathrm{VH}}|$ in MLG and BLG devices for $L \approx 1.5$ and 0.7 μm, respectively; $B = 40$ mT. Shaded (omitted) areas: Cyclotron diameter becomes comparable to the device width (*21*).

Figure 2b shows the $T$ dependence of $\alpha_{\mathrm{VH}}$ obtained using data such as those in Fig. 2a. Above liquid nitrogen $T$, where the hydrodynamic regime becomes fully developed (*3*), the viscous HE reaches 20% of the classical HE and has the opposite sign (Fig. 2b). $|\alpha_{\mathrm{VH}}|$ decreases with increasing $T$ and eventually disappears below our noise level above room $T$. Figs. 2d and e detail the observed behavior by plotting $\alpha_{\mathrm{VH}}(T, n)$ for MLG and BLG. The maps are somewhat different because of different viscosities of the two graphene systems (*3*, *22*, *23*) but show similar trends as functions of $n$ and $T$. We have also studied how $\alpha_{\mathrm{VH}}$ depends on $L$ and found that it rapidly decreases with $L$, practically disappearing if the voltage probe is placed further than $\sim 2$ μm from the current-injecting contact (Fig. 2c). The latter dependence highlights the importance of the vicinity geometry to detect viscous effects and explains why the Hall viscosity did not reveal itself in previous experimental studies of graphene.

Let us now turn to theory of electron magnetohydrodynamics. In the linear-response and steady-state regimes, two-dimensional viscous transport in the presence of a perpendicular field $B$ (in the $z$ direction) is described by the Navier-Stokes equation

$$\frac{\sigma_0}{ne}\nabla\phi(\boldsymbol{r}) = (1 - D_\nu^2\nabla^2)\boldsymbol{v}(\boldsymbol{r}) + \omega_c\tau(1 + D_H^2\nabla^2)\boldsymbol{v}(\boldsymbol{r}) \times \boldsymbol{z}, \tag{1}$$

in conjunction with the continuity equation and no-slip boundary conditions (*12*). Here, $\boldsymbol{v}(\boldsymbol{r})$ is the local fluid velocity, $\omega_c = eB/m$ the cyclotron frequency for electrons with the effective mass $m$, $\sigma_0 = ne^2\tau/m$ the Drude conductivity and $\tau$ the transport time with respect to momentum-non-conserving collisions such as, e.g., scattering on phonons. The right-hand side of Eq. (1) contains two terms. The first describes the electric current and viscous friction parameterized through the diffusion constant, $D_\nu = \sqrt{\nu\tau}$. The second term arises from the Lorentz force $\boldsymbol{F}_L = -(\omega_c m)\,\boldsymbol{v}(\boldsymbol{r}) \times \boldsymbol{z}$ and its viscous counterpart that depends on $\nu_H$ and is parameterized through another diffusion constant, $D_H = \sqrt{\nu_H/\omega_c}$. Note that the Hall friction acts against $\boldsymbol{F}_L$ which also means that $\nu_H$ does not perform any work on the electron fluid and, therefore, it is a dissipationless coefficient.

For the half-plane geometry (fair approximation for our devices) and relatively close to the injection point, the magnetohydrodynamic equations can be solved analytically (*12*) yielding $\phi(\boldsymbol{r})$ shown in Figs. 1a-d (fig. S4 provides examples of the potential and current maps calculated taking into account the finite device width). The Hall contribution to the vicinity resistance can be written as

$$R_A = -\sigma_0^{-1}\xi\left(\frac{L}{D_\nu}\right)\frac{\nu_H}{\nu}, \tag{2}$$

where $\xi(x) = [L_1(x) - I_1(x)]/2x$, and $L_1(x)$ and $I_1(x)$ are the modified Struve and Bessel functions, respectively (*12*). The function $\xi(L/D_\nu)$ decreases monotonically with $L$, behaving as $D_\nu/\pi L$ for $L \gg D_\nu$. The $L$ dependence expected for our devices is plotted in Fig. 2c, showing qualitative agreement with the experiment, even without taking into account the finite width (~0.3 μm) of our current and voltage probes.

The measured $R_A(B)$ such as shown in Fig. 2a can be used to determine $\nu_H$. To this end, we rewrite Eq. (2) as $\nu_H = -R_A\sigma_0\nu/\xi(L/\sqrt{\nu\tau})$ where $\sigma_0$ and $\tau$ are found from standard longitudinal resistivity measurements (*3*, *9*). The longitudinal viscosity $\nu(B)$ can be approximated using the semiclassical expression (*10*, *12*, *24*) $\nu(B) = \nu_0\frac{B_0^2}{B^2+B_0^2}$ where $\nu_0$ is the kinematic viscosity in zero $B$, and $B_0 = \hbar v_F k_F/(8e\nu_0)$ is the characteristic magnetic field expressed through the Fermi wave number $k_F$, the Fermi velocity $v_F$, and the reduced Planck constant $\hbar$. For the reported range of $n$ and $T$, $B_0$ is much larger than the fields in our experiments. Accordingly, we can assume $\nu \approx \nu_0$ in Eq. (2) and use $\nu_0$ from the experiments (*9*). Employing the above protocol, it is straightforward to calculate $\nu_H$ and its $B$ and $T$ dependences. Examples are shown in Figs. 3a-b. One can see that the Hall viscosity is linear in $B$ and rapidly decreases with increasing $T$.

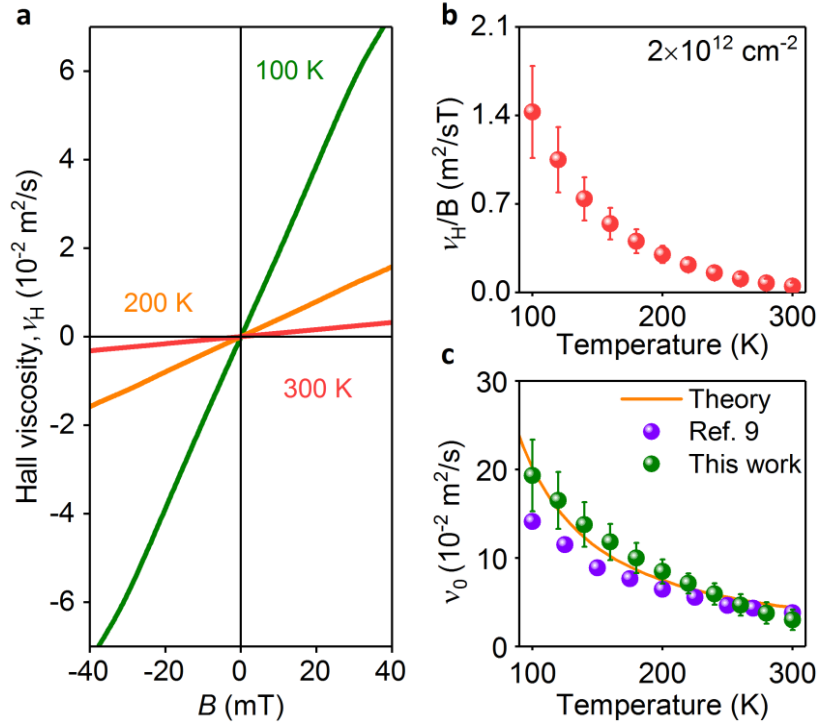

FIG. 3: **Hall viscosity in graphene.** (**a**) $\nu_H$ extracted using $R_A(B)$ for $L = 1$ μm. (**b**) $\nu_H/B$ as a function of $T$. (**c**) Zero-field viscosity $\nu_0$ found from our $R_A(B)$ measurements (green). Solid curve: Theory (*25*). Purple symbols: Previous experiments (*9*). For all the panels: MLG at $n = 2 \times 10^{12}$ cm$^{-2}$. No fitting parameters were used for the theory curve in (c). Error bars in (b) and (c) represent the scatter for measurements using different $L$.

For consistency, we crosschecked the above analysis against the results obtained previously (*9*) for zero-field viscosity $\nu_0$. To this end, we note that the field dependence of $R_A = R_A[\nu_H(B), \nu(B)]$ originates from changes in both longitudinal and Hall viscosities. The full formula for $\nu(B)$ is given above whereas the same semiclassical consideration (*10, 12*) for the Hall viscosity yields $\nu_H(B) = \nu_0 \frac{BB_0}{B^2 + B_0^2}$. This allows us to redefine the anomalous Hall contribution as $R_A[\nu_0, B]$ and calculate $\nu_0$ from the measured $R_A(B)$ dependences such as in Fig. 2a. Figure 3c compares $\nu_0$ extracted using this procedure with the values found independently in Ref. (*9*). The figure flaunts good agreement between the two analyses and with the viscosity expected theoretically (*25*). As for BLG, its Hall viscosity was evaluated using the same approach and reported in fig. S3n.

Finally, the above hydrodynamic description is also consistent with the large negative magnetoresistance observed in graphene devices above liquid-nitrogen $T$ (*18*). Similar magnetoresistance was reported in other high-quality 2D systems at elevated $T$ (*6, 26, 27*) but only recently it was realized that the anomaly could be caused by viscous effects (*6, 10, 11, 27*). In our devices, the high-$T$ negative magnetoresistance can be described accurately, without any fitting parameters (fig. S5), using the same $\nu_0$ as found experimentally (Fig. 3c). This unambiguously shows that the anomalous magnetoresistance is a hydrodynamic phenomenon, too. Another related effect we observed (*18*) is that $\nu_H$ suppresses Hall resistivity if it is measured locally, using contacts close to the current injector (e.g., in the geometry $R_{35,12}$ described above). The suppressed Hall effect is detailed in fig. S6 and provides an alternative way to determine Hall viscosity (*18*). Such measurements yield exactly the same $\nu_H$, despite being less clear-cut experimentally and overshadowed by a large contribution from the classical HE (*18*).

To conclude, Hall viscosity strongly affects electron transport, especially in the regions where the electric current is non-uniform. By probing local potentials in such regions we have succeeded to measure graphene's Hall viscosity and its $B, n, T$ dependences. It would be interesting to expand these studies into the quantum Hall effect regime, which attracts considerable theory interest (*28–30*) but no experimental procedure has so far been suggested to probe this regime.

# Supplementary Information

## #1 Device fabrication and characterization

Our devices were made of monolayer and bilayer graphene crystals. In the fabrication process, we employed the same procedures as reported previously (*8*, *31*, *32*). In brief, graphene was first encapsulated between relatively thick (~50 nm) hBN crystals using the standard dry-peel technique (*31*). The resulting stack was then deposited on top of an oxidized Si wafer (290 nm of $SiO_2$), which served as a back gate electrode. After this, the van der Waals assembly was patterned using electron beam lithography to define contact regions. Reactive ion etching was implemented to selectively remove the areas unprotected by a lithographic mask, which resulted in trenches for depositing long electrical leads. Metal contacts to graphene were then made by evaporating 3 nm of Cr and 80 nm of Au. Our devices were often endowed with a top gate. To this end, another round of electron-beam lithography was used to prepare a thin metallic mask which defined a multiterminal Hall bar and, at the same time, served as a top gate electrode. Subsequent plasma etching translated the shape of this mask into graphene. For Hall bar devices without a top gate, their shape was defined in the same manner but using a PMMA mask instead of a metal one. Optical micrographs of the resulting devices are shown in Fig. 1e of the main text and figs. S1a-b. The studied devices were $2 - 4$ μm in width $W$ and up to 20 μm in length and had several closely spaced voltage probes allowing us to study the local response at various distances $L$ from the current injector.

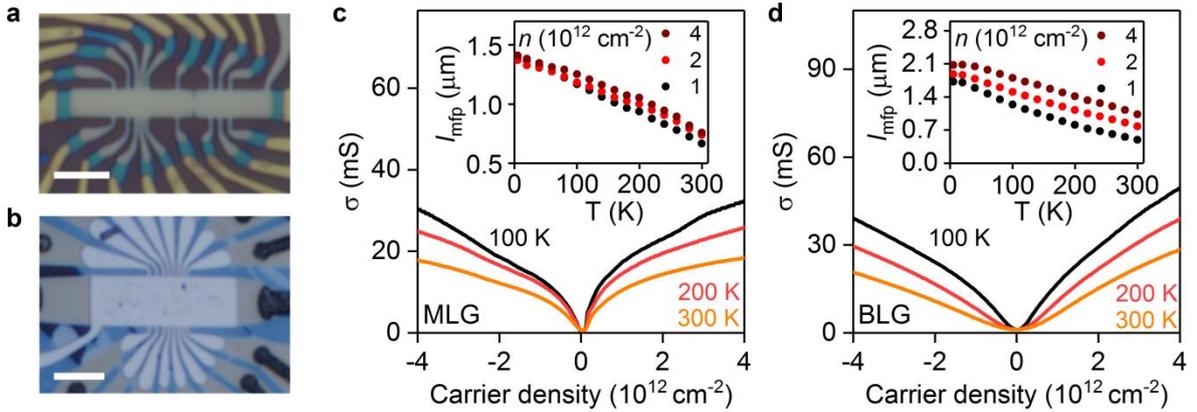

FIG. S1: Further examples of the studied graphene devices and their behavior. (**a** and **b**) Optical micrographs of bilayer (a) and monolayer (b) devices. Graphene Hall bars repeat the shape of the top gate (white bright areas). Numerous metallic leads terminated with quasi-one-dimensional contacts to graphene are seen in a dull yellow color in (a). Other colors correspond to different etching depths of the hBN-graphene-hBN stack. Scale bars: 4 μm in both (a) and (b). (**c**) Examples of $\sigma(n)$ at a given $T$ for the device shown in Fig. 1e of the main text. Insert: Mean free path $l_{\mathrm{mfp}}$ as a function of $T$ for different $n$. (**d**) Same as (c) but for BLG. Data are acquired at zero displacement field between the two graphene layers.

After fabrication, the devices were characterized using the standard four-terminal geometry. This involved measurements of their conductivity $\sigma$ as a function of carrier density $n$ and temperature $T$ (figs. S1c and d for MLG and BLG, respectively). One can see a typical field-effect behavior for high-quality graphene. In particular, the conductivity minimal at the charge neutrality point steeply rises with increasing $n$. The charge carrier mobility $\mu$ was calculated using the Drude formula, $\mu = \sigma/ne$, and for typical $n = 10^{12}$ cm$^{-2}$ exceeded 150,000 cm$^2$/Vs at 4 K and remained around 100,000 cm$^2$/Vs at room temperature for most of our MLG and BLG devices. These values translate into the elastic mean free path $l_{\mathrm{mfp}} = \mu\hbar/e(\pi n)^{1/2}$ of about $0.5 - 2$ μm (insets of figs. S1c-d).

## #2 Negative vicinity resistance

For consistency with the previous reports (*3*, *9*), we have checked that all our devices exhibited the key signature associated with electron hydrodynamics. It is the negative sign of the four-probe resistance measured close to the current injector (see the inset of fig. S2 for the schematic of the measurements). As expected, we observe a non-monotonic temperature dependence of the vicinity resistance $R_{\mathrm{v}} = V/I$ such that $R_{\mathrm{v}}$ first decreases with increasing $T$, changes its sign to become negative above $T \approx 50 - 100$ K, goes through a minimum at about $100 - 200$ K and then starts to grow, becoming positive again. This non-monotonic behavior and negative $R_{\mathrm{v}}$ above liquid-nitrogen $T$ were observed for all our devices reported in this work, in agreement with the previous experiments (*3*, *9*). The effect is explained by viscous entrainment of the electron fluid by the injected current, which produces negative lobes in the electrical potential $\phi$ in a vicinity of the injector (*3*, *14*, *15*) (see Fig. 1 of the main text).

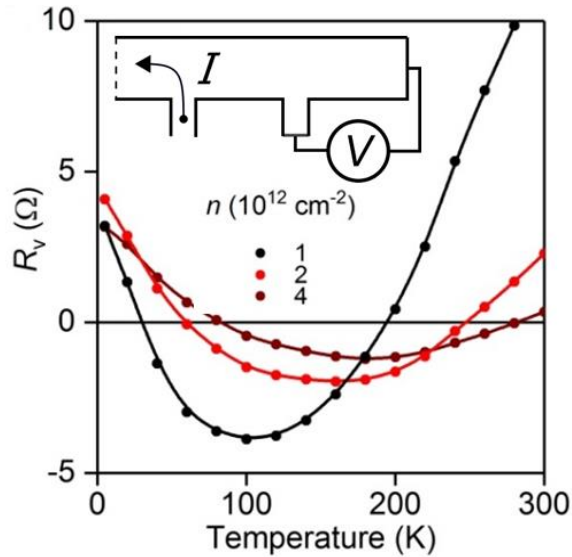

FIG. S2: Vicinity resistance. Examples of $R_{\mathrm{v}}$ as a function of $T$ for different $n$ away from the charge neutrality where doping is non-uniform. The data are for BLG. The distance $L$ between the current contact and vicinity-voltage probe is $\sim 0.7$ μm. Inset: Measurements'

geometry. Current $I$ is injected from a narrow lead into a wide graphene channel and the local potential $V$ is measured using a probe closest to the injector (*3*, *14*).

### #3 Viscous Hall effect in bilayer graphene

To emphasize that the antisymmetric-in-$B$ contribution to $R_v$ is a general feature of hydrodynamic transport in graphene and was observed for different devices and using different contacts, fig. S3a shows another example of $R_A(B)$. In this case, measurements were carried out for a BLG device with $\mu \approx 160,000$ cm$^2$/Vs at liquid-helium $T$. One can see that the curves in fig. 3a are qualitatively similar to those in Fig. 2a of the main text for MLG. The plotted dependences are linear in $B$ over the entire field interval up to 40 mT used in this work. The linearity corroborates the use of the viscous Hall coefficient $\alpha_{VH}$ to describe the behavior of $R_A(B)$ as discussed in the main text.

Furthermore, using the same analysis as presented in the main text, we have extracted the Hall viscosity of BLG's electron fluid. To this end, we used the experimental dependences $R_A(B)$ in fig. S3a and the equation $\nu_H = -R_A \sigma_0 \nu / \xi (L/\sqrt{\nu\tau})$ provided in the main text and valid for BLG, too. Again, $\sigma_0$ and $\tau$ were found from the standard resistivity measurements whereas $\nu \approx \nu_0$ was taken from measurements of the vicinity resistance in zero $B$ (*3*). The results for $\nu_H(B)$ at two characteristic $T$ are shown in fig. S3b. One can see the Hall viscosity for BLG's electron fluid behaves similar to that in MLG and exhibits comparable values (cf. Fig. 3a-b of the main text).

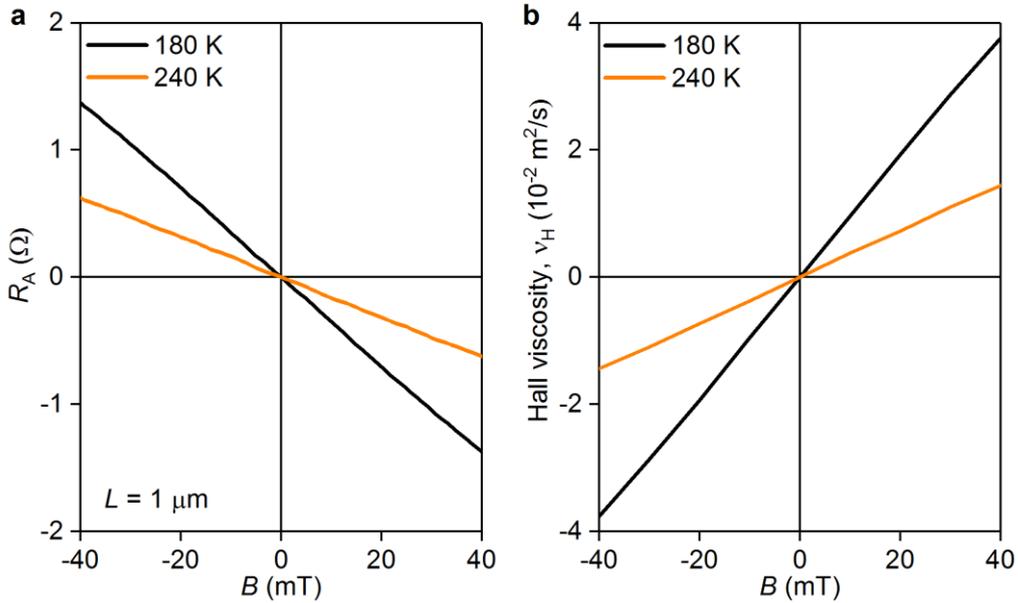

FIG. S3: Hall viscosity in BLG. (**a**) Examples of $R_A(B)$ at the given $L$ in the $T$ range where electron hydrodynamics dominates. (**b**) $\nu_H$ obtained from the data in (A) and using $\nu \approx \nu_0$ from Ref. (*3*). $n = 2 \times 10^{12}$ cm$^{-2}$ for both panels.

**#4 Current and potential distribution in a Hall bar**

In the main text, Figs. 1a-b show the distribution of the electrical potential $\varphi(\mathbf{r})$ near a point contact injecting electric current into the half-plane. Those distributions could be found analytically by solving the Navier-Stokes equation [Eq. (1) of the main text]. However, our devices have the finite width $W$ and, to find changes in $\varphi(\mathbf{r})$ and the current density distribution $\mathbf{J}(\mathbf{r})$ for the experimental geometry, we have solved the Navier-Stokes equation numerically using the approach described in Refs. (*12*, *14*). Examples of such calculations are given in fig. S4 that shows $\varphi(\mathbf{r})$ and $\mathbf{J}(\mathbf{r})$ in zero and finite $B$. At zero $B$ (fig. S4a), a viscous flow results in an anomalous region of negative $\varphi$ in the vicinity of the current-injecting contact (red arrow). This negative lobe is accompanied by a whirlpool of the electric current (see the current lines in fig. S4), in agreement with the previous reports (*3*, *14*, *15*, *17*). A finite magnetic field (fig. S4b) causes considerable changes in $\varphi(\mathbf{r})$ such that the negative lobe expands and the current whirlpool becomes larger. In general, the behavior of $\varphi(\mathbf{r})$ with increasing $B$ is consistent with the analytical solutions shown in Fig. 1 of the main text. Although the confinement creates current whirlpools, it does not change qualitatively the anomalous voltage response (negative lobe) that appears near the contact to the right from the current injector in fig. S4.

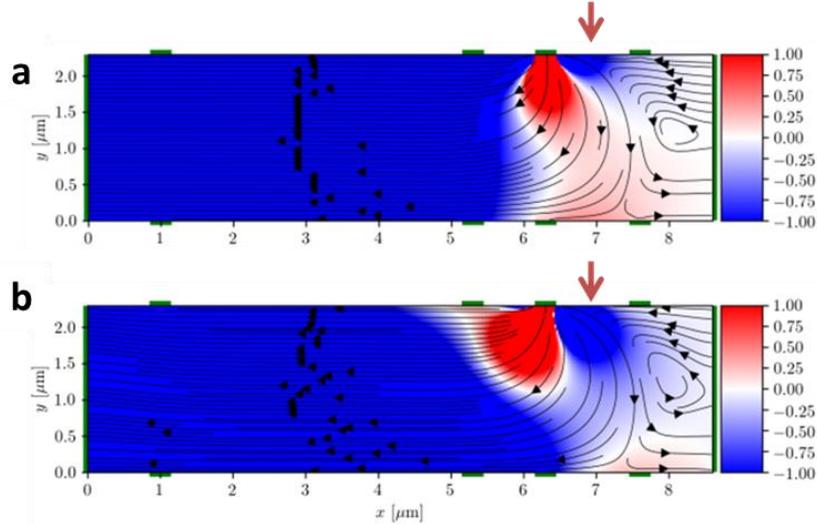

FIG. S4: Current and voltage distributions in the Hall bars used in our experiments. (**a**) Calculated $\mathbf{J}(\mathbf{r})$ and $\varphi(\mathbf{r})$ in the vicinity of the current injector in zero magnetic field. (**b**) Same for a finite $B$ such as $\nu_H = 2\nu$. The green bars indicate voltage contacts and $D_\nu = 1$ μm for both panels. Current streamlines are shown by black curves with arrows. The thick red arrows point to the regions of the negative potential caused by the viscous flow. In (b), the trivial contribution from the classical Hall effect (see Fig. 1c of the main text) is subtracted to emphasize changes in the negative lobe in $\varphi$ which are seen to the right from the injector in both (a) and (b). The color-coded potential is given in units of $I/\sigma_0$.

**#5 Anomalous negative magnetoresistance**

High-mobility two-dimensional systems are known to exhibit large negative magnetoresistance (MR) at cryogenic temperatures (*26, 33, 34*). This MR is routinely explained in terms of skipping orbits that appear in a magnetic field and suppress backscattering (*11, 21, 33*). This results in a rapid increase in conductance (decrease in resistance) above a certain magnetic field $B$ in which the cyclotron diameter becomes comparable to the device width $W$. Unsurprisingly, our devices also exhibit a large negative MR at low $T$. However, we also observed a similarly large MR at $T$ above 200 K. Under these conditions, $l_{\mathrm{mfp}}$ is notably smaller than $W$ (see fig. S1), and the description in terms of ballistic transport and skipping orbits becomes objectionable. This invites alternative explanations. One such explanation has recently been put forward in terms of electron magnetohydrodynamics (*10, 11*), which seems quite appropriate for our case of high-quality graphene above liquid-nitrogen $T$.

Following arguments similar to those in Refs. (*10, 11*), we argue below that the negative MR observed in our devices above 200 K (fig. S5) is fully consistent with the same magnetohydrodynamics description that is used to explain other transport anomalies in graphene at elevated $T$. To this end, we solve the Navier-Stokes equation for a graphene strip and, for no-slip boundary conditions, obtain the analytical expression (*14*)

$$\rho = \frac{\sigma_0^{-1}}{(1 - 2(D_\nu/W)\tanh[W/2D_\nu])},$$ (S1)

This equation yields that the resistivity $\rho$ should decrease with decreasing $\nu$ and $D_\nu = \sqrt{\nu\tau}$. Physically, this is a result of the decreasing friction between the electron fluid and sample boundaries. Next, we use the semiclassical prediction (*10, 12, 24*)

$$\nu(B) = \nu_0 \frac{B_0^2}{B^2 + B_0^2}.$$ (S2).

The decrease of $\nu$ with increasing $B$, given by Eq. (S2), can be qualitatively understood as follows (*24*). The kinematic viscosity is a measure of momentum transfer between fluid elements and, therefore, is proportional to the electron-electron mean free path $l_{\mathrm{ee}}$. Magnetic field shortens $l_{\mathrm{ee}}$, limiting it to the cyclotron diameter. The latter is inversely proportional to $B$ and, hence, $\nu$ should drop with increasing $B$. Combining Eqs. (S1 and S2), it is straightforward to understand why magnetic field suppresses $\rho$ in the hydrodynamic regime.

Figure S5 plots the normalized magnetoresistance $\mathrm{MR}(B) = [\rho(B) - \rho(0)]/\rho(0)$. The solid curves are our measurements for MLG and BLG devices whereas the dashed curves are the theoretical expectations for $\mathrm{MR}(B)$ using Eq. (S1) in which $\nu$ is defined through Eq. (S2) and $\nu_0$ is taken from the experiment (Fig. 3c). The plots show good agreement between the experiment and theory, especially at higher $T$ where the hydrodynamic description is expected to hold better. Importantly, no fitting parameters are used in these plots, which is a good indicator that the negative MR is indeed caused by the hydrodynamic behavior.

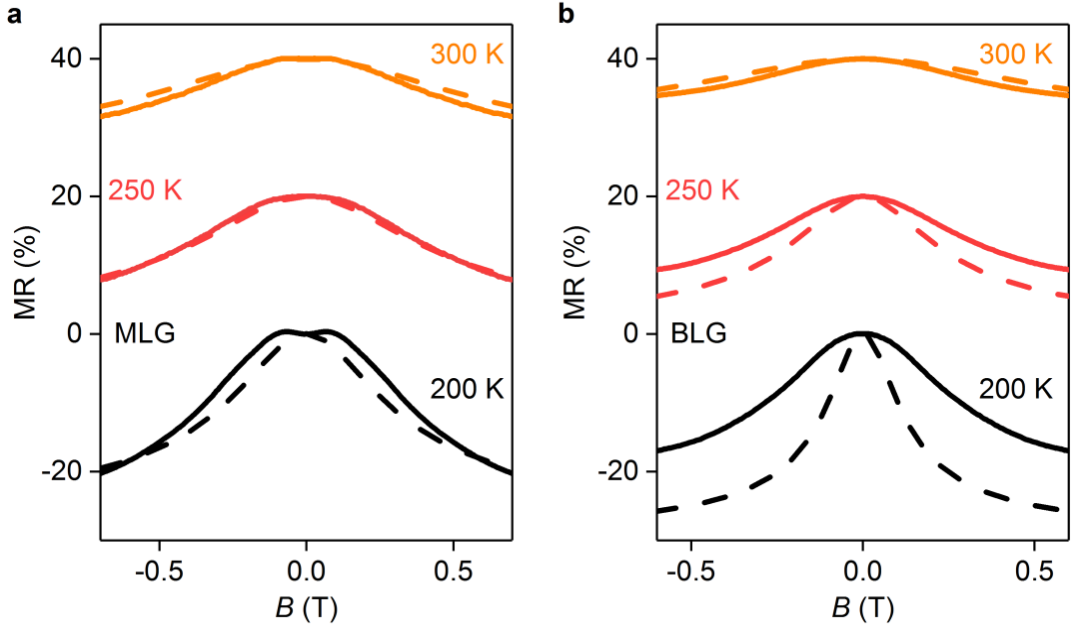

FIG. S5: Negative magnetoresistance in the hydrodynamic regime. (**a**) MLG at $n = 2 \times 10^{12}$ cm$^{-2}$. (**b**) Same for BLG. In both cases, $W = 2$ μm. The solid curves are experimental data. Dashed: Theory using Eqs. (S1-S2) and experimental values of $\nu_0$ such as in Fig. 3c. For clarity, the curves for different $T$ are shifted by 20 %.

## #6 Local suppression of the Hall effect

There exists another measurement geometry that allows measurements of $\nu_H$. As discussed in the main text, Hall viscosity suppresses the Hall resistance. Because viscous effects are prominent at micrometer ($\sim D_\nu$) distances from the region of the inhomogeneous current flow created by the current injector (Fig. 1a-d of the main text), one can also expect violations of the Hall effect if it is measured locally, close to the injector region. Indeed, we have found that the local Hall resistance $R_{\mathrm{LH}}$ measured as shown in the inset to fig. S6a is quite different in value from the classical Hall resistance, $R_{\mathrm{H}} = B/ne$. The Hall resistance measured in both conventional and local geometries was found linear in $B$ at all $T$ but local measurements yielded a value suppressed by as much as 30% with respect to the classical HE, if the distance between the current contact and two voltage probes was $\sim 1$ μm (fig. S6a). The observed linearity in $B$ allows us to define the Hall coefficients $\alpha_{\mathrm{H}} = R_{\mathrm{H}}/B(ne)^{-1}$ and $\alpha_{\mathrm{LH}} = R_{\mathrm{LH}}/B(ne)^{-1}$ (fig. S6b). For a metal with a single type of charge carriers, the classical HE is described by $\alpha_{\mathrm{H}} \equiv 1$, and our measurements in the conventional Hall geometry show no deviations from the unity over the entire $T$ range (fig. S6b). In stark contrast, the local Hall coefficient, also shown in fig. S6b, strongly depends on $T$ so that $\alpha_{\mathrm{LH}}$ becomes nearly twice smaller at 100 K than the standard value of 1. Using the magnetohydrodynamic analysis reported in the main text, we can show that these deviations are also caused by Hall viscosity. Indeed, it is straightforward to find that $R_{\mathrm{LH}}$ can be written as

$$R_{LH} = R_H + 2R_A, \qquad (S3)$$

where $R_A$ is a function of $\nu_H$ as discussed in the main text. We emphasize that $R_H$ and $R_A$ have opposites signs so that the absolute value of $R_{LH}$ is always smaller that $R_H$ and the difference increases with increasing $\nu_H$ and decreasing $L$. Furthermore, as follows from Eq. (S3), simultaneous measurements of $R_{LH}$ and $R_H$ allow us to find $\nu_H$, alternatively to the protocol discussed in the main text. Results of the analysis using Eq. (S3) are plotted in fig. S6c and compared with the $\nu_H(T)$ dependence found from the vicinity measurements (Fig. 3b). One can see that the two approaches yield excellent agreement.

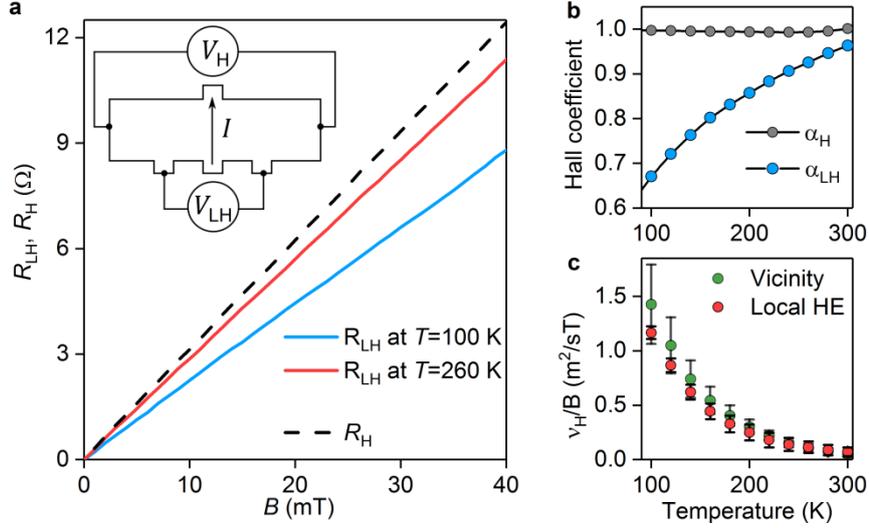

FIG. S6: Local Hall effect. (**a**) Local Hall resistance $R_{LH}$ at two $T$ within the hydrodynamic regime (solid lines). Dashed: Conventional measurements yield the classical result $R_H = B/ne$. Inset: Schematics of measuring $R_H$ and $R_{LH}$. (**b**) Hall coefficients obtained from the slopes of $R_H(B)$ and $R_{LH}(B)$ such as in (a). MLG at $n = 2 \times 10^{12}$ cm$^{-2}$ and $L \approx 1$ μm. (**c**) $\nu_H/B$ as a function of $T$ extracted from the local Hall and vicinity resistance measurements. The latter data are taken from Fig. 3b of the main text. Error bars: Statistics using measurements at different $L$.